\documentclass[aps,prd,groupedaddress,showpacs]{revtex4}
\usepackage{amsmath}

\begin{document}

\title{A wide family of singularity-free cosmological models}

\author{L. Fern\'andez-Jambrina}
\email[]{lfernandez@etsin.upm.es}
\homepage[]{http://debin.etsin.upm.es/lfj.htm}
\affiliation{E.T.S.I. Navales\\ Universidad Polit\'ecnica de Madrid \\ 
Arco de la Victoria s/n \\ E-28040 Madrid, Spain}

\author{L.M. Gonz\'alez-Romero}
\affiliation{Departamento de F\'\i sica Te\'orica II\\ 
Facultad de Ciencias F\'\i sicas \\
Universidad Complutense de Madrid \\ 
Avenida Complutense s/n \\ E-28040 Madrid, Spain}
\email[]{mgromero@fis.sim.ucm.es}

\date{\today}

\begin{abstract}
In this paper a family of non-singular cylindrical perfect fluid 
cosmologies is derived. The equation of state corresponds to a stiff 
fluid. The family depends on two independent functions under very 
simple conditions. A sufficient condition for geodesic completeness is 
provided.\end{abstract}


\maketitle

\section{Introduction}

After the discovery of the first regular perfect fluid cosmological model by 
Senovilla \cite{seno}, one of the major questions that that spacetime posed was 
to determine how generic this lack of singularities was. For instance, 
in the Ruiz-Senovilla \cite{ruiz} barotropic family with equation of state $p= 
\gamma\mu$, $0< \gamma< 1$, the regular models were those 
corresponding to a radiation fluid, $\gamma=1/3$. These solutions are 
the separatrix between those models with singular pressure and  energy density, 
$\gamma>1/3$, and those who just had singular Weyl curvature, $\gamma<1/3$ \cite{esc}. Therefore regular 
models would be a zero-measure set in this family.

Our aim in this paper is to determine an infinite family of regular 
cosmological models that need not be so restrictive as the 
Ruiz-Senovilla family and therefore may indicate that regular models cannot be 
neglected in the set of solutions of Einstein equations.

We shall focus on stiff perfect fluids, since they are simple enough 
to allow almost complete integration of Einstein equations and 
thereby constitute an excellent arena for checking hypotheses. We 
shall show that under very simple restrictions regular solutions appear.

The second section of this paper will be devoted to deriving solutions 
of the stiff fluid Einstein equations in a convenient manner for our 
purposes. In the third section geodesic completeness of the solutions 
shall be imposed and the restrictions derived from this assumption 
shall be expressed as a sufficient condition.

\section{Stiff fluid cosmologies}

We shall restrict to spacetimes endowed with an Abelian orthogonally 
transitive group of isometries, $G_{2}$, acting on timelike surfaces, 
since this is the framework where regular cosmological models have so 
far appeared in. We shall further require that the Killing fields be 
mutually orthogonal. Adapting the coordinates to these fields we 
shall write them as $\{\partial_{z},\partial_{\phi}\}$. Under these assumptions we can write the metric 
for the spacetime in a convenient coordinate chart \cite{kramer}, $\{t,r,z,\phi\}$,
\begin{equation}
ds^2=e^{2K}(-dt^2+dr^2)+e^{-2U}dz^2+\rho^2e^{2U}d\phi^2,\label{metric}
\end{equation}
 which has been chosen as isotropic on the non-ignorable 
coordinates, $t,r$. This is a priori no restriction and may  
always be achieved since every 2-metric admits an isotropic 
parametrization. The metric is written in terms of three functions, 
$K$, $U$ and $\rho$ that depend only on $t$ and $r$.

We may interpret the isometry group as cylindrical symmetry in the 
spacetime provided we have a regular axis where the norm of the 
angular Killing field vanishes. We shall come back to this issue later 
on. The range of the coordinates will be then the usual one for cylindrical symmetry,
\begin{equation}
    -\infty<t,z<\infty,\ 0<r<\infty,\ 0<\phi<2\pi.
\end{equation}

The matter content of the spacetime will be a perfect fluid of energy 
density $\mu$, pressure $p$ and 4-velocity $u$. The 
energy-momentum tensor will be then,
\begin{equation}
    T^{\mu\nu}=\mu u^{\mu}u^\nu+p\,(g^{\mu\nu}+u^\mu u^\nu),\qquad 
    0\le\mu,\nu\le 3,\qquad u^\mu u_{\mu}=-1.
\end{equation} 

For a stiff fluid, $\mu=p$. We shall write down the Einstein 
equations, $R_{\mu\nu}-R\,g_{\mu\nu}/2=T_{\mu\nu}$, in a comoving system of 
coordinates for the perfect fluid, that is, $u=e^{-K}\partial_{t}$. After some simplications the 
equations read,
\begin{subequations}
\label{einstein1} 
\begin{eqnarray}
U_{tt}-U_{rr}+\frac{1}{\rho}(U_{t}\rho_{t}-U_{r}\rho_{r})=0,\label{U1}
\\
\rho_{tt}-\rho_{rr}=0,\label{rho1}
\\
K_{t}\rho_{r}+K_{r}\rho_{t}=\rho_{tr}+U_{t}\rho_{r}+U_{r}\rho_{t}+2\rho 
U_{t}U_{r},\label{Kr1}
\\
K_{t}\rho_{t}+K_{r}\rho_{r}=\frac{\rho_{tt}+\rho_{rr}}{2}+U_{t}\rho_{t}+U_{r}\rho_{r}
+\rho(U_{t}^2+U_{r}^2)+p\rho e^{2K},\label{Kt1}
\\
K_{rr}-K_{tt}+\frac{U_{r}\rho_{r}-U_{t}\rho_{t}}{\rho}+U_{r}^2-U_{t}^2=pe^{2K}, \label{nu1}
\\
K_{r}+\frac{p_{r}}{2p}=0,\label{p1}
\\
K_{t}+\frac{\rho_{t}}{\rho}+\frac{p_{t}}{2p}=0,\label{mu1}
\end{eqnarray}
\end{subequations}
where the last pair of equations are just Euler and continuity equations for the 
perfect fluid.

Since every regular cosmological model in the literature has a $\rho$ with spacelike 
gradient, we shall impose as an ansatz that $\mathrm{grad}\,\rho$ be 
orthogonal to the velocity of the fluid, $u$. 

Under this assumption, $\rho$ will be a function of $r$ only. But 
then (\ref{rho1}) requires that $\rho$ be a linear function of $r$. 
After rescaling the coordinates we can take $\rho=r$ and the whole 
system of equations becomes rather simple,

\begin{subequations}
\label{einstein} 
\begin{eqnarray}
U_{tt}-U_{rr}-\frac{U_{r}}{r}=0,\label{U2}
\\
K_{t}=U_{t}+2r U_{t}U_{r},\label{Kr2}
\\
K_{r}=U_{r}+r(U_{t}^2+U_{r}^2)+pr e^{2K},\label{Kt2}
\\
K_{rr}-K_{tt}+\frac{U_{r}}{r}+U_{r}^2-U_{t}^2=pe^{2K}, \label{nu2}
\\
K_{r}+\frac{p_{r}}{2p}=0,\label{p2}
\\
K_{t}+\frac{p_{t}}{2p}=0.\label{mu2}
\end{eqnarray}
\end{subequations}

The energy-momentum conservation equations can be integrated, 
\begin{equation}
    p=\alpha e^{-2K},
\end{equation} with $\alpha=\mathrm{const}.>0$ 
 and equation (\ref{nu2}) is a 
consequence of the others. We are left then with a two-dimensional 
reduced wave equation in polar coordinates 
without source term (\ref{U2}) and a quadrature for $K$,
\begin{subequations}
\label{quadrature} 
\begin{eqnarray}
K_{t}=U_{t}+2r U_{t}U_{r},\label{Kr}
\\
K_{r}=U_{r}+r(U_{t}^2+U_{r}^2)+\alpha r,\label{Kt}
\end{eqnarray}
\end{subequations}that can be integrated 
after providing a solution to the wave equation. The integrability 
condition for the quadrature (\ref{quadrature}) is namely the wave 
equation, so the whole problem reduces to solving it.

Following \cite{john}, for instance, the solution to the Cauchy 
problem for the wave equation in the plane can be constructed from 
the $3D$ solution by ignoring the  third variable, 
\begin{equation}
U(x,y,t)=\frac{1}{2\pi}\int_{0}^{2\pi}d\phi\int_0^tdR 
R\frac{g(x+R\cos\phi,y+R\sin\phi)}{\sqrt{t^2-R^2}}+\frac{1}{2\pi}\frac{\partial}{\partial 
t}\int_{0}^{2\pi}d\phi\int_0^tdR 
R\frac{f(x+R\cos\phi,y+R\sin\phi)}{\sqrt{t^2-R^2}},
\end{equation} 
for initial data $U(x,y,0)=f(x,y)$, $U_{t}(x,y,0)=g(x,y)$. Note that 
the $2D$ wave 
equation does not satisfy Huygens' principle and therefore the domain 
of dependence is a circle, not just a circumference.

Our problem is even easier since we do not have dependence on the 
polar angle. Therefore we just have to impose circular symmetry on the 
initial data $U(r,0)=f(r)$, $U_{t}(r,0)=g(r)$. The time coordinate may be removed from the integration limits 
by an appropriate scaling, $R=t\tau$,
\begin{equation}
U(r,t)=\frac{1}{2\pi}\int_{0}^{2\pi}d\phi\int_0^1d\tau 
 \frac{\tau}{\sqrt{1-\tau^2}}\left\{tg(v)+f(v)+tf'(v) 
 \frac{t\tau^2+r\tau\cos\phi }{v}
\right\},\end{equation}
where $v=\sqrt{r^2+t^2\tau^2+2rt\tau\cos\phi}$, choosing the origin 
of the polar angle at the angle for $(x,y)$.  This expression is valid 
for all values of $t$.

For instance, the non-singular spacetime in \cite{leo} is generated by 

\begin{displaymath}
    f(x)=\frac{\beta}{2}x^2,\qquad g(x)= 0,\qquad\beta>0.
\end{displaymath}

It is clear now that we are just integrating the functions $f,g$ on a 
finite interval of $r$ and that the integrals are well-defined 
provided $g,f,f'$ are continuous. The singularity at $\tau=1$ is 
harmless under such conditions.

The solution does not 
share the class of differentiability of the initial data because of 
the derivative term. We need at least $f\in C^3([0,\infty))$, $g\in 
C^2([0,\infty))$ in order to have $U,K\in C^2([0,\infty))$ and a well-defined Riemann tensor.

Surprisingly there is no need to impose cylindrical symmetry on the 
solution, since 
we have a regular axis at $r=0$ provided that $f,g$ are regular there 
and we have already required it. According to \cite{kramer}, the axis 
is regular if
\begin{equation}
    \lim_{r\to0}\frac{\langle 
        \textrm{grad}\,\Delta,\textrm{grad}\,\Delta\rangle}{4\Delta}=
	e^{2(U-K)}|_{r=0}=1, 
        \quad 
        \Delta=\langle\partial_{\phi},\partial_{\phi}\rangle=r^2e^{2U},
    \label{axis}
\end{equation}
therefore we need that $K(0,t)=U(0,t)$ for every value of $t$. But at 
the axis the equations (\ref{Kr},\ref{Kt}) that determine $K$ are rather simple,
\begin{equation}\label{Kaxis}
    K_{r}=U_{r},\qquad K_{t}=U_{t},
\end{equation} and therefore $K(0,t)=U(0,t)+\mathrm{const}.$ and the 
condition of regularity at the axis is fulfilled by either taking the 
constant of integration equal to zero or conveniently rescaling the angular 
coordinate.

Note that this requirement of regularity excludes a 
timelike gradient of $\rho$ in the vicinity of the axis.

\section{Geodesic completeness}

The metric that is obtained after integrating the system 
(\ref{Kr},\ref{Kt}) has regular components in the whole spacetime, but 
this not suffices in order to have a non-singular spacetime. We shall 
consider that a spacetime is regular \cite{HE} if it is causally 
geodesically complete, that is, if every causal geodesic may be 
extended to all values of its affine parametrization.

This means analysing the geodesic equations for diagonal cylindrically 
symmetric spacetimes. This has been done in \cite{manolo}. Those 
results can be summarized as follows:

\begin{description}
\item[Theorem:] A cylindrically symmetric diagonal metric in the form (\ref{metric}) with $C^2$
metric functions  $f,g,\rho$ is future causally geodesically complete 
provided that along causal geodesics:
\begin{enumerate}
\item For large values of $t$ and increasing $r$, 
\begin{enumerate}
    \item $K_{r}+K_{t}\ge 0$, and either $K_{r}\ge 0$ or 
    $|K_{r}|\lesssim K_{r}+K_{t}$.
    \item $(K+U)_{r}+(K+U)_{t}\ge 0$, and either $(K+U)_{r}\ge 0$ 
    or $|(K+U)_{r}|\lesssim (K+U)_{r}+(K+U)_{t}$.
    \item $(K-U-\ln\rho)_{r}+(K-U-\ln\rho)_{t}\ge 0$, and either 
    $(K-U-\ln\rho)_{r}\ge 0$  or $|(K-U-\ln\rho)_{r}|\lesssim 
    (K-U-\ln\rho)_{r}+(K-U-\ln\rho)_{t}$.
\end{enumerate}

\item \label{tt} For large values of  $t$, constant $b$ exist such that 
$\left.\begin{array}{c}K(t,r)-U(t,r)\\2\,K(t,r)\\K(t,r)+U(t,r)+\ln\rho(t,r)
\end{array}\right\}\ge-\ln|t|+b.$

\end{enumerate}\end{description}

A similar result can be stated for past-pointing geodesics just 
reversing the sign of the time derivatives in condition 1.

Note that we have omitted the condition for non-radial geodesics with 
decreasing $r$ in \cite{manolo}, since according to (97) in that 
reference, non-radial causal geodesics would reach 
$r=0$ for finite $t$, that is, before $t$ becomes singular, 
contradicting the fact that geodesics should be singular there. 
Therefore the axis cannot be reached by geodesics with non-zero 
angular momentum.

All we have to do now in order to have a geodesically complete model 
is to check whether conditions 1 and 2 are satisfied.

We first show that the conditions on the derivatives are always
fulfilled for a stiff fluid model:

\begin{enumerate}
    \item  
    \begin{enumerate}
        \item  According to (\ref{Kr},\ref{Kt}) we have, 
        \begin{displaymath}
            K_{t}+K_{r}=U_{t}+U_{r}+r (U_{t}+U_{r})^2+\alpha r.
        \end{displaymath} Several possibilities are open: When 
        $U_{t}+U_{r}$ is positive, $K_{t}+K_{r}$ is positive. If 
        $U_{t}+U_{r}$ is negative and $|U_{t}+U_{r}|\ge 1$, the 
        quadratic term is larger and $K_{t}+K_{r}$ is again positive 
        for large values of $r$. Finally, if $U_{t}+U_{r}$ is negative 
	and $|U_{t}+U_{r}|\le 1$, it is the pressure term $\alpha r$ which 
	overcomes the negative term for large $r$. 
	
	The same sort of reasoning is valid to conclude that the radial derivative,
        \begin{displaymath}
            K_{r}=U_{r}+r (U_{t}^2+U_{r}^2)+\alpha r.
        \end{displaymath} is positive for large values of $r$.
    
        \item We can apply the same argument to $K+U$,
        \begin{displaymath}
            (K+U)_{t}+(K+U)_{r}=2(U_{t}+U_{r})+r (U_{t}+U_{r})^2+\alpha r,
        \end{displaymath} in order to show that these derivatives and 
        $(K+U)_{r}$ are positive for large values of the radial 
        coordinate.
	
        \item Finally, the third condition is always satisfied,
	\begin{displaymath}
            (K-U-\ln\rho)_{t}+(K-U-\ln\rho)_{r}=
	    r (U_{t}+U_{r})^2-\frac{1}{r}+\alpha r\ge0,
        \end{displaymath}
	\begin{displaymath}
            (K-U-\ln\rho)_{r}=
	    r (U_{t}^2+U_{r}^2)-\frac{1}{r}+\alpha r\ge0,
        \end{displaymath} for increasing $r$.
    
    \end{enumerate}
    
    Past-pointing geodesics are treated analogously without any 
    additional problem, since reversing the sign of time derivatives 
    does not alter the positivity of the quadratic terms. 
    
    Consequently only condition 2. yields a restriction.

    \item As we shall see, this condition amounts to study $U$ at the 
    axis for large values of the time coordinate:
    
    \begin{enumerate}
        \item  This condition is trivial since $(K-U)|_{r=0}=0$, 
	\begin{displaymath}
	    \big(K-U\big)(t,r)=\int_{0}^{r}dr'\,(K-U)_{r}(t,r')=
	    \int_{0}^{r}dr'\,\left\{r'\big(U_{t}^2(t,r')+U_{r}^2(t,r')\big)+
	    \alpha r\right\}>0.
	\end{displaymath}
    
        \item  The previous reasoning for ruling out 
	singularities for decreasing radius leaves us with two possibilities: 
        increasing radius and constant radius, $r=0$. Since, according to 
	(\ref{Kaxis})
    \begin{displaymath}
        K(r,t)=U(0,t)+\int_{0}^rdr'\,K_{r}(r',t),
    \end{displaymath} and we have already checked that $K_{r}$ is positive for 
    large $t$ and increasing $r$, we just have to study the term 
    $U|_{r=0}$, that is, we have reduced the problem to analysing the 
    behaviour of $K$ at the axis 
    for large values of $t$.
    
        \item  Similarly, \begin{displaymath}
        \big(K+U\big)(r,t)+\ln r=2U(0,t)+\ln r +\int_{0}^rdr'\,\big(K+U\big)(r',t),
    \end{displaymath} and we have already checked the positivity of 
    $K_{r}+U_{r}$ as in the previous condition. The logarithmic term 
    does not mean a problem for increasing radius. Again we are left 
    with controlling the behaviour of the $U$ term.
    \end{enumerate}
\end{enumerate}

Summarizing our results so far, in order to have a causally 
geodesically complete spacetime we just have to require that 
$U|_{r=0}$ does not decrease faster than a negative logarithm for 
large values of the absolute value of the time coordinate. The condition on the solution of the Cauchy 
problem for the wave equation at the axis becomes a bit simpler,

\begin{equation}
K(0,t)=U(0,t)=\int_0^1d\tau 
 \frac{\tau}{\sqrt{1-\tau^2}}\left\{tg(|t|\tau)+f(|t|\tau)+|t|\tau 
 f'(|t|\tau) 
\right\}\ge -\frac{1}{2}\ln |t|+b,\label{req}\end{equation} 
since the dependence on  the polar angle is lost.

This bound can be attained, for example, by the initial data,
\begin{displaymath}
    f(r)=\frac{|a|-1}{2}\ln r,\qquad g(r)=\frac{a}{\pi}\frac{\ln r}{r},
\end{displaymath}
since for this choice of functions the solution to the Cauchy problem for 
the wave equation is, 
\begin{displaymath}
    U(0,t)=\frac{1}{2}\big(|a|+a\,\textrm{sign}\,t-1\big)\ln|t|
    +\frac{1}{2}\big(|a|-a\,\textrm{sign}\,t-1\big)\ln2=
    \left\{
    \begin{array}{ll}
       \frac{1}{2}\big(2|a|-1\big)\ln|t|-\ln2 & \textrm{sign}\,at>0  
       \\ \\
        \frac{1}{2}\big(2|a|-1\big)\ln2-\frac{1}{2}\ln|t| & \textrm{sign}\,at<0
    \end{array}
    \right..
\end{displaymath}

The behaviour of the terms in (\ref{req}) is rather different. The 
term, $U_{f}$, dependent on $f$, the initial value of $U$, is even in the time 
coordinate, as it is to be expected when the initial time derivative of 
$U$ is zero. On the contrary, the term $U_{g}$, dependent on $g$, is odd in $t$. 

This means that if $U_f$ satisfies (\ref{req}) for 
 positive time, it is automatically satisfied for 
negative time. On the contrary, if  $U_g$  satisfies 
(\ref{req}) for positive time, it is only satisfied for 
negative time if it is also satisfied by $U_{-g}$ for positive time. 
Therefore three different possibilities are open depending on the 
value of 
\begin{equation}
    \lim_{r\to \infty}\frac{f(r)+rf'(r)}{rg(r)}.\label{lim}
\end{equation}

\begin{itemize}
    \item  If (\ref{lim}) is infinite, we need $U_{f}(t)>-\frac{1}{2}\ln|t|+b$ 
    for large values of $t$ in order to have geodesic completeness. 
    This means that $f(r)+rf'(r)>-\frac{1}{2}\ln r +k$ for large $r$.

    \item  If (\ref{lim}) is zero, we need 
    $|U_{g}(t)|<\frac{1}{2}\ln|t|+b$ 
    for large values of $t$ in order to have geodesic completeness. 
    This means that $|g(r)|<\frac{1}{\pi r}\ln r+k$ for large $r$.

    \item  If (\ref{lim}) is finite, then $U_{f}$ and $U_{g}$ are of 
    the same order for large values of $t$ or of $-t$ and geodesic 
    completeness will depend on the value of the limit.
\end{itemize}

Note that values of the integrand close to $\tau=0$ do not influence 
the result for large values of $t$, since we may split the integral in two terms,
\begin{displaymath}
    U(0,t)=\int_{0}^{x_{c}/t}+\int_{x_{c}/t}^1,
\end{displaymath}
and the first one is bounded and negligible for large $t$. Therefore 
the main contribution to $U$ comes from the second term, which must 
fulfill the required asymptotic behaviour.

\section{Examples}

A simple and wide family of functions that satisfy (\ref{req}) can 
be written in terms of polynomials. Consider

\begin{equation}
    f(r)=\sum_{i=0}^na_{i}r^i,\qquad g(r)=\sum_{i=0}^mb_{i}r^i.
\end{equation}
 If $n,m$ are even 
numbers, $U$ can be analytically integrated in terms of polynomials. For our purposes we just 
require $U$ at the axis, which can be integrated for a larger set of 
functions. Since $U$ is linear in $f$ and $g$, we may analyse the monomials 
independently. For $f(r)=r^n$, $g(r)=r^m$ we obtain,
\begin{equation}
    U_{f}(t)=\frac{n!!}{(n-1)!!}\left(\frac{\pi}{2}\right)^{(1+(-1)^{n+1})/2}|t|^n,\qquad 
    U_{g}(t)=\frac{m!!}{(m+1)!!}\left(\frac{\pi}{2}\right)^{(1+(-1)^{m+1})/2}|t|^mt.
\end{equation}
These expressions are valid even for $n,m=-1$ taking $(-1)!!:=1$, 
although they may be not very practical.

According to (\ref{req}) we have two different possibilities for obtaining a singularity-free 
model:

\begin{itemize}
    \item  If $f,g$ are polynomials in $r$ respectively of degree 
    $n,m$ and $n>m+1$, we have a non-singular model if $a_{n}$  is positive.

    \item  If $f,g$ are polynomials in $r$ respectively of degree 
    $n,n-1$, $U_{f}$ and $U_{g}$ at the axis are polynomials of 
    degree $n$ and we have a non-singular model if $U_{f}$ dominates 
    over $U_{g}$. This happens if the leading term of $U_{f}$ is 
    greater than the one of $U_{g}$, that is,
    
 \begin{equation}
 \left(\frac{n!!}{(n-1)!!}\right)^2\left(\frac{2}{\pi}\right)^{(-1)^n}
 \frac{a_{n}}{|b_{n-1}|}>1.\end{equation}   
 Using Stirling's formula for approximation of factorials, an easy and safe bound 
 would be
  \begin{equation}
 \left(n+\frac{1}{2}\right)\,a_{n}>|b_{n-1}|. \label{term}\end{equation} 
 
 This family of non-singular cosmological models is large indeed, as 
 it can be seen by restricting the range to a finite dimensional space 
 of polynomial functions: if we consider the space of functions $U$ 
 for which $U|_{r=0}$ is a polynomial of degree equal or lower than 
 $n$, the subset of singularity-free models comprises an open set, 
 according to (\ref{term}).
 
 This result can be generalized, since 
  \begin{equation}
      \int_{0}^1d\tau \,\frac{\tau^{p+1}}{\sqrt{1-\tau^2}}=
 \frac{\sqrt{\pi}}{2}\frac{\Gamma((p+2)/2)}{\Gamma((p+3)/2)}\end{equation} 
 allows integration for every real value of the exponent $p$. Therefore, 
 for $f(r)=r^p$, $g(r)=r^q$, we obtain,
 \begin{equation}
     U_{f}(t)=
 \sqrt{\pi}\frac{\Gamma((p+2)/2)}{\Gamma((p+1)/2)}|t|^{p},\qquad
      U_{g}(t)=
 \frac{\sqrt{\pi}}{2}\frac{\Gamma((q+2)/2)}{\Gamma((q+3)/2)}|t|^{q}t,\end{equation}
 which allow generalization of the geodesic completeness requirements which have been previously 
 derived for polynomial functions to linear combinations of powers 
 of $r$. Additionally 
 one has just  to care about the class of differentiability of $U$, 
 which demands that $p\ge2$, $q>2$.
  
\end{itemize}

\section{Conclusions}

We have analysed a wide family of stiff perfect fluid cosmological models with 
cylindrical symmetry. The issue of causal geodesic completeness has 
been reduced to just the behaviour at the axis of the initial value 
problem for 
a sourceless 2D-wave equation, which is the only one left after 
simplifying Einstein equations. A sufficient condition for geodesic 
completeness is provided, which is very easy to check and to 
implement. The case of polynomial initial data has been discussed and 
allows a fairly large set of non-singular cosmological models. We 
think that this set is wide enough to preclude considering non-singular 
models as isolated points in a space of cosmological models.

The role of pressure in these models is obviously determinant, since 
stiff perfect fluids are a limit case for energy conditions, 
corresponding to a sound velocity equal to that of light. On the 
contrary, dust perfect fluids are always singular according to 
Raychaudhuri equation. Intermediate cases remain open for discussion, 
even though partial results have been obtained \cite{esc}.

It is interesting to notice that nonseparability of the models in 
these coordinates is fundamental for geodesic completeness. In 
\cite{agnew} separable cosmologies were studied and \emph{none} of 
them was found to be regular.

\begin{acknowledgments}
The present work has been supported by Direcci\'on General de
Ense\~nanza Superior Project PB98-0772. The authors wish to thank
 F.J. Chinea,  F. Navarro-L\'erida and  M.J. Pareja 
for valuable discussions. 
\end{acknowledgments}

\end{document}